\documentclass{aa}
\usepackage{caption}
\usepackage{graphicx}
\usepackage{txfonts}
\usepackage{hyperref}
\usepackage{blindtext}
\usepackage{multirow}
\hypersetup{colorlinks=true,linkcolor=blue,citecolor=blue,filecolor=blue,urlcolor=blue,}
\newcommand{\kms} {\TextOrMath{\,km s$^{-1}$}{\mathrm{\,km s}^{-1}}}
\newcommand{\msol}{\TextOrMath{\,M$_\odot$}{\mathrm{\,M}_\odot}}

\newcommand{\about} {$\sim$}

\begin{document} 

\title{Binary neutron star merger offsets from their host galaxies}
\subtitle{II. Short-duration gamma-ray bursts}
\titlerunning{Binary neutron star merger offsets from their host galaxy II}

\author{
N. Gaspari\inst{1}
\and
A.~J. Levan\inst{1,2}
\and
A.~A. Chrimes\inst{3,1}
\and
A.~E. Nugent\inst{4,5}
}

\institute{
Department of Astrophysics/IMAPP, Radboud University, P.O. Box 9010, 6500 GL Nijmegen, the Netherlands\\
\email{nicola.gaspari@live.it}
\and  
Department of Physics, University of Warwick, Coventry CV4 7AL, UK
\and 
European Space Agency (ESA), European Space Research and Technology Centre (ESTEC), Keplerlaan 1, 2201 AZ Noordwijk, the Netherlands
\and
Center for Astrophysics | Harvard \& Smithsonian, 60 Garden St. Cambridge, MA 02138, USA
\and
Center for Interdisciplinary Exploration and Research in Astrophysics (CIERA) and Department of Physics and Astronomy, Northwestern University, Evanston, IL 60208, USA
}

\date{Received ---; accepted ---}
 
\abstract
{The mergers of binary neutron stars (BNSs) and neutron star-black holes (NSBHs) binaries have long been linked to short-duration gamma-ray bursts (SGRBs). 
However, despite their stellar progenitors, SGRBs are often found outside the stellar light of the host galaxy.
This is commonly attributed to supernova kicks, which displace the SGRB progenitors from the original stellar population.}
{Our goal is to use stellar population synthesis models to reproduce and interpret the observed offsets of a statistical sample of SGRBs, using realistic galactic models based on the observed host properties.}
{We derive the host galaxy potentials from the observed properties on a case-by-case basis, and simulate the galactic trajectories of synthetic BNSs and NSBHs from the \texttt{BPASS} code using three different kick prescriptions. We compare predicted and observed offsets to investigate the impact of velocity kicks, host galaxy types, and host association criteria.}
{The results confirm that the locations of the SGRB population are consistent with the expectations of kicked BNS or BHNS progenitors, implying that such mergers are the dominant (if not only) progenitor system. 
Predictions for NSBHs provide a significantly worse fit compared to BNSs, while we find no significant difference when comparing different kick prescriptions. 
For late-type hosts, we find the best agreement when including hosts with a probability of chance alignment $P_\mathrm{ch}$ up to 20\%, while lower $P_\mathrm{ch}$ thresholds lead us to overestimate SGRB offsets. 
We argue that $P_\mathrm{ch}$ is biased against viable hosts at the largest offsets, and suggest the use of less conservative $P_\mathrm{ch}$ thresholds for late-type hosts.
For early-type hosts, the predictions underestimate SGRB offsets in a few cases regardless of the $P_\mathrm{ch}$ threshold applied. 
We argue that this is likely due to the models missing galaxy evolution, or spurious host associations.}
{}

\keywords{stars: neutron -- gamma-ray burst: general -- gravitational waves}

\maketitle


\section{Introduction}

Short-duration gamma-ray bursts \citep[SGRBs;][]{kouveliotou1993} are one of the manifestations of binary neutron star (BNS) mergers, along with kilonovae and afterglows in several bands. 
This connection has been established through several different SGRB observables, including their redshift distribution, the lack of coinciding supernovae (SNe), the demographics of their host galaxies, and most recently, the coincident detection of GRB~170817 and the binary neutron star merger GW~170817 \citep[for a review see][]{nakar2007,lee2007,berger2014,2017ApJ...848L..12A}.
Despite BNSs being considered the main progenitor, it remains unclear whether SGRBs are representative of BNS mergers, since not all BNS mergers might produce a SGRB \citep{rastinejad2022,troja2022,sarin2022,salafia2022,levan2024,yang2024} and not all SGRBs might be produced by a BNS merger \citep{qin1998,levan2006,metzger2008,troja2008,gompertz2020}. 
Furthermore, SGRBs are oftentimes found near a galaxy but outside its stellar light \citep[e.g. Fig.~2 of][]{fong2022}, suggesting that SGRB locations do not trace stellar light despite having stellar progenitors. 

Since SGRBs do not always spatially coincide with a galaxy, identifying their host galaxies is a non-trivial task. A common approach to this problem is to compute the probability of chance alignment $P_\mathrm{ch}$ of each galaxy around the SGRB location, and identify the one with the lowest probability as the host \citep{bloom2002}. Lower $P_\mathrm{ch}$ values correspond to higher likelihoods of correctly assigning the host.  However, in $\sim20-30\%$ of cases, no host can be confidently associated or multiple galaxies have the same non-negligible $P_\mathrm{ch}$, deeming the event hostless \citep{berger2010,fong2013b,tunnicliffe2013,oconnor2022}.
For the hosts with a strong association, SGRBs are located at offsets that extend well beyond the host stellar light, and hence they do not trace the host stellar light \citep{fong2013,oconnor2022,fong2022}. Despite this apparent discrepancy, the offset distribution is consistent with the predictions for BNS mergers \citep{narayan1992,pzy1998,bloom1999,fryer1999,bulik1999,bloom2002,perna2002,vosstauris2003,belczynski2006,church2011,mandhai2022,gaspari2024a}, since BNS systems can travel with high systemic velocities given the two velocity kicks the system receives at each NS formation from the SN explosion \citep[e.g.][and references therein]{andrews2019}.
There are however competing explanations for the highest offsets and the hostless bursts, even within the BNS formation scenario. For instance, large merger offsets can be achieved by BNSs that formed within the stellar light but received a kick large enough to escape the host \citep[e.g.][]{zemp2009,kelley2010,behroozi2014,beniamini2016,wiggins2018,zevin2020}, as well as by BNSs that formed in the host outskirts \citep[either in a globular cluster or in a faint and extended stellar halo, e.g.][]{salvaterra2010,church2011,bae2014,perets2021}. Hostless bursts instead could be produced by BNS that escaped the host, or by BNS merging at a redshift high enough for the host galaxy to be faint and hence undetected \citep{berger2010,tunnicliffe2013,mandhai2022,oconnor2022}.

Understanding the offset distribution of SGRBs is therefore crucial in order to identify the host galaxies, discriminate between formation scenarios, and constrain the physical models for compact object mergers. In the literature there are already several works that synthesised populations of BNSs and neutron star-black hole (NSBH) binaries, and simulated their galactic trajectories in order to study the merger locations. However, most of them either analysed only a few extreme cases of galactic models, such as potentials with very different masses, or different star formation histories \citep[SFHs;][]{bloom1999,fryer1999,bulik1999,bloom2002,perna2002,vosstauris2003,belczynski2006,salvaterra2010}, or used potentials and SFHs from cosmological simulations \citep{zemp2009,kelley2010,behroozi2014,wiggins2018,perna2022,mandhai2022}. Only a few works used galactic models that are based on the observed properties of real SGRB hosts \citep{abbott2017,zevin2020,gaspari2024b}, and to date only \cite{church2011} used this approach with a statistical population of hosts. As the number of observed SGRB hosts has grown significantly in the last decade, this work aims to expand that of \cite{church2011} by modelling and analysing in a systematic way the locations of an entire SGRB population, using galactic models that reproduce case by case the observed host properties.

The paper is structured as follows. In Section~\ref{sec:2} we describe the sample of SGRB host galaxies, the method to model the galactic potentials, and the simulations of BNS trajectories within the potentials. In Section~\ref{sec:3} we present the predicted BNS merger locations and compare them to the observed SGRB locations, before we summarise and conclude in Section~\ref{sec:4}. Throughout the paper, magnitudes are corrected for Galactic extinction along the line of sight \citep{schlegel1998,schlafly2011} and reported in the AB system \citep{oke1982}. We adopt a $\Lambda$CDM cosmology with $H_0=67.66$ \kms Mpc$^{-1}$, $\Omega_\Lambda=0.69$, and $\Omega_\text{m}=0.31$ \citep{planck2020}.


\section{Methods}\label{sec:2}

\subsection{Host galaxies sample}

In this work, we use the sample of SGRBs and respective host galaxies assembled and characterised by \cite{fong2022} and \cite{nugent2022} (hereafter, the BRIGHT\footnote{\url{http://bright.ciera.northwestern.edu/}} sample). 
The sample consists of 90 short GRBs observed mostly by NASA’s Neil Gehrels Swift Observatory \citep{gehrels2004}, and characterised by a localisation with $\leq5"$ uncertainty, a clear sight line outside of crowded fields or high Galactic extinction regions, and a probability of chance alignment $P_\mathrm{ch}\leq20\%$ \citep{bloom2002}. 
The sample contains $\sim60\%$ of the Swift population. 
Our analysis is limited to the subsample of 70 hosts that have a spectral energy distribution (SED) model, all of which are characterised by having a detection in at least 3 photometric bands. 
The subsample contains $\sim77\%$ of the BRIGHT sample, and $\sim46\%$ of the Swift population. 
We discuss the possible bias introduced by the $P_\mathrm{ch}$ association criterium in Sect.~\ref{sec:3.3}, while the effects of other selection criteria are discussed in detail in \cite{fong2022} and \cite{nugent2022}.
The subsample properties are listed in Table~\ref{tab:1}.

The host galaxies are classified by \cite{nugent2022} as star-forming, transitioning, and quiescent, following the specific star-formation rate prescription of \cite{tacchella2022}. 
In the following, we refer to the star-forming galaxies as late (L) type, and to the transitioning and quiescient galaxies as early (E) type. 
The hosts are also divided into three classes based on $P_\textrm{ch}$, with the Gold (G) sample containing hosts with $P_\textrm{ch}\leq 0.02$, the Silver (S) sample cointaining those with $0.02<P_\textrm{ch}\leq 0.10$, and the Bronze (B) sample containing those with $0.10<P_\textrm{ch}\leq 0.20$. 
For 50 out of 70 hosts we have spectroscopic redshifts, while for the remaining 20 we use the photometric redshifts obtained by \cite{nugent2022} from the SED fitting. 
For 19 hosts we also have an inferred Sersic profile for the surface brightness, collected from different works in the literature. 
The projected offsets between bursts and host galaxies are taken from \cite{fong2022}, and the respective uncertainties combine the GRB localisation uncertainty, the galaxy centroid uncertainty, and the astrometric tie uncertainty.

\begin{table*}
    \centering
    \caption{
    Short GRB host sample used in this work. 
    The columns list the GRB name, the host galaxy type, the host redshift, the projected offset, the rest-frame total magnitude in the $B$-band, the stellar mass, the mass-averaged age, the scale age, the observed half-light radius, the observed Sersic index, the half-light radius from the mass-size relation, and the references for the listed values.}
    \label{tab:1}
    \bgroup
    \def\arraystretch{1.29} 
    \begin{tabular}{lccccccccccc}  
        \hline\hline
        \noalign{\smallskip}
GRB	& Host type & $z$ & Offset	& $M_B$ & $\log M_\star/\msol$ & $t_\mathrm{m}$ & $\tau$ & $R_\mathrm{half}$ & $n$ & $R_\mathrm{half}^\mathrm{ms}$ & Ref. \\
& & & [kpc] & & & [Gyr] & [Gyr] & [kpc] & & [kpc] & \\
        \noalign{\smallskip}
        \hline
        \noalign{\medskip}
\multicolumn{12}{c}{Gold sample}\\
        \noalign{\smallskip}
050509B	& Early	& 0.22\phantom{0}    & 55.19$\pm$12.43	& -22.10	& 11.46	& 8.84	& 1.07	& 20.98	& 5.60	& $6.28_{-0.70}^{+7.76}$	& 1,13	\\
050709	& Late	& 0.16\phantom{0}    & 3.760$\pm$0.056	& -17.63	& 8.55	& 0.57	& 0.10	& 1.75	& 0.60	& $2.10_{-0.92}^{+1.53}$	& 1,13	\\
050724A	& Early	& 0.26\phantom{0}    & 2.74$\pm$0.08	& -20.42	& 11.05	& 8.20	& 1.07	& 4.00	& 2.90	& $5.30_{-1.68}^{+5.08}$	& 1,13	\\
051221A	& Late	& 0.55\phantom{0}    & 2.08$\pm$0.19	& -19.91	& 9.31	& 0.49	& 0.14	& 2.49	& 0.90	& $2.79_{-1.22}^{+1.81}$	& 1,13	\\
060614	& Late	& 0.12\phantom{0}    & 0.70$\pm$0.79	& -15.57	& 7.85	& 0.76	& 0.14	& ---	& ---	& $1.43_{-0.62}^{+1.20}$	& 2,13	\\
060801	& Late	& 1.13\phantom{0}    & 10.25$\pm$10.92	& -21.19	& 9.12	& 0.13	& 0.48	& ---	& ---	& $2.19_{-1.00}^{+1.58}$	& 3,13	\\
061006	& Late	& 0.46\phantom{0}    & 1.39$\pm$0.29	& -18.28	& 9.37	& 4.27	& 1.21	& 3.67	& 0.70	& $3.30_{-1.45}^{+1.99}$	& 1,13	\\
061210	& Late	& 0.41\phantom{0}    & 15.51$\pm$14.36	& -19.41	& 9.49	& 0.66	& 0.11	& 2.18	& 1.03	& $3.53_{-1.54}^{+2.07}$	& 4,13	\\
070429B	& Late	& 0.90\phantom{0}    & 6.00$\pm$13.44	& -21.54	& 10.44	& 0.43	& 0.11	& 5.08	& 2.15	& $4.81_{-2.08}^{+2.38}$	& 4,13	\\
070714B	& Late	& 0.92\phantom{0}    & 12.33$\pm$0.87	& -19.74	& 9.37	& 1.63	& 2.07	& 2.20	& 1.18	& $2.89_{-1.26}^{+1.85}$	& 4,13	\\
070724A	& Late	& 0.46\phantom{0}    & 5.52$\pm$0.18	& -20.76	& 9.81	& 0.27	& 0.10	& 3.64	& 0.92	& $4.13_{-1.80}^{+2.27}$	& 4,13	\\
070809	& Early	& 0.47\phantom{0}    & 34.11$\pm$2.75	& -21.60	& 10.82	& 0.84	& 0.11	& 3.38	& 3.38	& $4.82_{-1.98}^{+3.94}$	& 4,13	\\
071227A	& Late	& 0.38\phantom{0}    & 14.74$\pm$0.26	& -19.86	& 10.79	& 1.79	& 0.73	& 4.72	& 1.05	& $6.98_{-2.76}^{+3.56}$	& 4,13	\\
090510	& Late	& 0.90\phantom{0}    & 10.51$\pm$2.92	& -21.02	& 9.75	& 0.45	& 0.10	& 7.27$^\S$	& 1.27$^\S$	& $3.63_{-1.54}^{+2.12}$	& 4,13	\\
100117A	& Late	& 0.91\phantom{0}    & 1.35$\pm$0.32	& -20.56	& 10.35	& 3.02	& 0.25	& 4.95$^\S$	& 0.55$^\S$	& $4.59_{-2.00}^{+2.39}$	& 3,4,13\\
100206A	& Late	& 0.41\phantom{0}    & 25.28$\pm$13.05	& -20.58	& 10.72	& 4.58	& 3.22	& ---	& ---	& $6.60_{-2.64}^{+3.41}$	& 2,13	\\
101224A	& Late	& 0.45\phantom{0}    & 12.75$\pm$13.51	& -19.65	& 9.17	& 0.46	& 0.14	& ---	& ---	& $2.96_{-1.29}^{+1.87}$	& 2,13	\\
120804A	& Late	& 1.05$^\star$       & 2.30$\pm$1.28	& -19.79	& 9.81	& 0.35	& 2.26	& ---	& ---	& $3.10_{-1.35}^{+1.91}$	& 2,13	\\
121226A	& Late	& 1.37$^\star$       & 2.31$\pm$9.15	& -21.43	& 9.46	& 0.12	& 2.00	& ---	& ---	& $2.61_{-1.16}^{+1.74}$	& 2,13	\\
130603B	& Late	& 0.36\phantom{0}    & 5.4$\pm$0.2	& -19.86	& 9.82	& 1.63	& 0.37	& 3.37$^\S$	& 1.29$^\S$	& $4.14_{-1.80}^{+2.27}$	& 4,13	\\
140129B	& Late	& 0.43\phantom{0}    & 1.76$\pm$1.76	& -17.85	& 9.33	& 1.65	& 0.37	& ---	& ---	& $3.23_{-1.41}^{+1.97}$	& 2,13	\\
140903A	& Late	& 0.35\phantom{0}    & 0.9$\pm$0.1	& -19.22	& 10.81	& 4.24	& 1.09	& ---	& ---	& $7.09_{-2.79}^{+3.61}$	& 2,13	\\
141212A	& Late	& 0.60\phantom{0}    & 18.75$\pm$12.29	& -19.84	& 9.71	& 2.37	& 1.39	& ---	& ---	& $3.54_{-1.51}^{+2.09}$	& 2,13	\\
150101B	& Early	& 0.13\phantom{0}    & 7.36$\pm$0.07	& -21.21	& 11.13	& 4.88	& 0.61	& 9.5	& 5.0	& $5.48_{-1.54}^{+5.53}$	& 2,5,13\\
150120A	& Late	& 0.46\phantom{0}    & 4.77$\pm$6.44	& -19.74	& 10.01	& 2.28	& 1.21	& ---	& ---	& $4.33_{-1.86}^{+2.37}$	& 2,13	\\
150728A	& Late	& 0.46\phantom{0}    & 7.52$\pm$20.29	& -20.45	& 9.35	& 0.15	& 0.10	& ---	& ---	& $3.27_{-1.43}^{+1.98}$	& 2,13	\\
160411A	& Late	& 0.81$^\star$       & 1.4$\pm$2.3	& -18.89	& 8.87	& 0.67	& 2.51	& ---	& ---	& $2.14_{-0.96}^{+1.54}$	& 2,13	\\
160525B	& Late	& 0.64$^\star$       & 5.50$\pm$7.38	& -18.44	& 8.04	& 0.14	& 2.21	& ---	& ---	& $1.30_{-0.61}^{+1.12}$	& 2,13	\\
170428A	& Late	& 0.45\phantom{0}    & 7.72$\pm$3.39	& -18.66	& 9.68	& 5.14	& 2.35	& ---	& ---	& $3.92_{-1.72}^{+2.19}$	& 2,13	\\
170728B	& Late	& 1.27\phantom{0}    & 6.76$\pm$2.06	& -21.99	& 9.87	& 0.41	& 0.34	& ---	& ---	& $3.17_{-1.38}^{+1.95}$	& 2,13	\\
170817A	& Early	& 0.01\phantom{0}    & 2.125$\pm$0.001	& -19.12	& 10.61	& 10.42	& 1.20	& 3.3	& 3.9	& $3.88_{-1.70}^{+2.80}$	& 6,13	\\
180418A	& Late	& 1.55$^\star$       & 1.30$\pm$0.32	& -21.20	& 9.83	& 0.56	& 1.67	& ---	& ---	& $2.73_{-1.22}^{+1.85}$	& 7,13	\\
180618A	& Late	& 0.52$^\star$       & 9.70$\pm$1.69	& -19.51	& 8.81	& 0.35	& 2.47	& ---	& ---	& $2.07_{-0.93}^{+1.51}$	& 2,13	\\
180727A	& Late	& 1.95$^\star$       & 2.56$\pm$5.13	& -20.48	& 9.22	& 0.54	& 1.75	& ---	& ---	& $2.11_{-0.95}^{+1.56}$	& 2,13	\\
181123B	& Late	& 1.75\phantom{0}    & 5.08$\pm$1.38	& -22.30	& 9.90	& 0.63	& 1.64	& ---	& ---	& $2.82_{-1.25}^{+1.87}$	& 8,13	\\
200219A	& Late	& 0.48$^\star$       & 8.30$\pm$5.28	& -20.93	& 10.74	& 3.53	& 2.34	& ---	& ---	& $6.71_{-2.67}^{+3.45}$	& 2,13	\\
200522A	& Late	& 0.55\phantom{0}    & 0.93$\pm$0.19	& -20.90	& 9.66	& 0.58	& 0.22	& 3.9	& 2.1	& $3.44_{-1.47}^{+2.06}$	& 9,13	\\
200907B	& Late	& 0.56$^\star$       & 2.41$\pm$7.78	& -19.37	& 9.35	& 0.89	& 1.44	& ---	& ---	& $2.86_{-1.24}^{+1.84}$	& 2,13	\\
210323A	& Late	& 0.73\phantom{0}    & 5.89$\pm$3.68	& -18.93	& 8.77	& 0.56	& 0.22	& ---	& ---	& $2.02_{-0.91}^{+1.48}$	& 2,13	\\
210726A	& Late	& 0.37$^\star$       & 12.26$\pm$2.22	& -15.63	& 7.84	& 1.06	& 2.11	& ---	& ---	& $1.42_{-0.62}^{+1.20}$	& 2,13	\\
211023B	& Late	& 0.86\phantom{0}    & 3.84$\pm$2.57	& -20.02	& 9.65	& 1.72	& 1.15	& ---	& ---	& $3.42_{-1.46}^{+2.05}$	& 2,13	\\
211211A	& Late	& 0.08\phantom{0}    & 7.920$\pm$0.029	& -17.38	& 8.84	& 2.53	& 0.86	& 1.64	& 1     & $2.46_{-1.08}^{+1.69}$	& 2,10,13\\
        \noalign{\smallskip}
        \hline
    \end{tabular}
    \egroup
\end{table*}

\begin{table*}
    \ContinuedFloat
    \centering
    \caption{Continued.}
    \bgroup
    \def\arraystretch{1.29} 
    \begin{tabular}{lccccccccccc}
        \hline\hline
        \noalign{\smallskip}
GRB	& Host type & $z$ & Offset	& $M_B$ & $\log\,M_\star/\mathrm{M}_\sun$ & $t_\mathrm{m}$ & $\tau$ & $R_\mathrm{half}$ & $n$ & $R_\mathrm{half}^\mathrm{ms}$ & Ref. \\
& & & [kpc] & [AB mag] & & [Gyr] & [Gyr] & [kpc] & & [kpc] & \\
        \noalign{\smallskip}
        \hline
        \noalign{\medskip}
\multicolumn{12}{c}{Silver sample}\\
        \noalign{\smallskip}
051210	& Late	& 2.58$^\star$       & 29.08$\pm$16.34	& -24.27	& 10.96	& 0.23	& 1.58	& ---	& ---	& $3.19_{-1.42}^{+2.19}$	& 1,13	\\
070729	& Late	& 0.52$^\star$       & 19.72$\pm$14.49	& -18.81	& 8.75	& 0.55	& 2.04	& ---	& ---	& $2.00_{-0.90}^{+1.47}$	& 2,13	\\
090515	& Early	& 0.40\phantom{0}    & 76.19$\pm$0.16	& -20.76	& 11.25	& 6.34	& 0.24	& 4.24$^\S$	& 2.95$^\S$	& $5.75_{-1.29}^{+6.27}$	& 4,13	\\
100625A	& Early	& 0.45\phantom{0}    & 2.63$\pm$6.77	& -18.84	& 9.70	& 3.56	& 0.21	& ---	& ---	& $1.90_{-0.81}^{+0.98}$	& 2,13	\\
101219A	& Late	& 0.72\phantom{0}    & 5.48$\pm$6.65	& -19.58	& 9.39	& 0.25	& 0.10	& ---	& ---	& $2.93_{-1.27}^{+1.87}$	& 2,13	\\
111117A	& Late	& 2.21\phantom{0}    & 10.52$\pm$1.68	& -22.65	& 9.63	& 0.19	& 1.30	& ---	& ---	& $2.13_{-0.92}^{+1.44}$	& 2,13	\\
120305A	& Early	& 0.22\phantom{0}    & 18.09$\pm$5.25	& -18.25	& 9.17	& 2.11	& 0.38	& ---	& ---	& $1.86_{-0.79}^{+1.03}$	& 2,13	\\
130515A	& Early	& 0.80$^\star$       & 61.22$\pm$13.77	& -22.09	& 10.29	& 0.78	& 0.10	& ---	& ---	& $1.77_{-0.67}^{+0.97}$	& 2,13	\\
130822A	& Early	& 0.15\phantom{0}    & 60.1$\pm$4.9	& -20.08	& 10.16	& 2.16	& 0.39	& ---	& ---	& $2.30_{-1.00}^{+1.32}$	& 2,13	\\
140930B	& Late	& 1.46\phantom{0}    & 9.62$\pm$4.3	& -21.18	& 9.45	& 0.57	& 2.35	& ---	& ---	& $2.59_{-1.16}^{+1.73}$	& 2,13	\\
150831A	& Late	& 1.18\phantom{0}    & 12.21$\pm$9.77	& -20.43	& 9.49	& 0.51	& 2.00	& ---	& ---	& $2.65_{-1.18}^{+1.75}$	& 2,13	\\
151229A	& Late	& 0.63$^\star$       & 8.16$\pm$6.05	& -17.27	& 8.79	& 1.78	& 1.37	& ---	& ---	& $2.04_{-0.92}^{+1.49}$	& 2,13	\\
160303A	& Late	& 1.01$^\star$       & 15.3$\pm$0.9	& -19.29	& 9.51	& 1.05	& 1.52	& ---	& ---	& $2.67_{-1.19}^{+1.76}$	& 2,13	\\
160624A	& Late	& 0.48\phantom{0}    & 9.63$\pm$6.24	& -19.85	& 9.74	& 1.19	& 0.42	& 6.8	& 1     & $4.05_{-1.77}^{+2.23}$	& 2,12,13\\
160821B	& Late	& 0.16\phantom{0}    & 15.74$\pm$0.03	& -19.26	& 9.24	& 0.58	& 0.12	& ---	& ---	& $3.07_{-1.35}^{+1.92}$	& 2,13	\\
161001A	& Late	& 0.67$^\star$       & 18.54$\pm$6.22	& -20.28	& 9.73	& 0.77	& 0.16	& ---	& ---	& $3.59_{-1.53}^{+2.11}$	& 2,13	\\
161104A	& Early	& 0.79\phantom{0}    & 1.66$\pm$16.60	& -20.41	& 10.23	& 2.26	& 0.28	& ---	& ---	& $1.70_{-0.66}^{+0.94}$	& 11,13	\\
170127B	& Late	& 2.21$^\star$       & 10.37$\pm$13.60	& -21.80	& 9.51	& 0.31	& 1.69	& ---	& ---	& $2.03_{-0.86}^{+1.41}$	& 2,13	\\
180805B	& Late	& 0.66\phantom{0}    & 24.30$\pm$7.49	& -20.17	& 9.34	& 0.50	& 0.22	& ---	& ---	& $2.84_{-1.24}^{+1.83}$	& 2,13	\\
191031D	& Late	& 1.93$^\star$       & 13.08$\pm$10.69	& -22.32	& 10.38	& 0.80	& 1.43	& ---	& ---	& $3.49_{-1.51}^{+2.01}$	& 2,13	\\
        \noalign{\smallskip}
        \hline
        \noalign{\medskip}
\multicolumn{12}{c}{Bronze sample}\\
        \noalign{\smallskip}
050813	& Late	& 0.72\phantom{0}    & 43.57$\pm$17.37	& -20.29	& 10.31	& 3.73	& 1.30	& ---	& ---	& $4.50_{-1.96}^{+2.39}$	& 2,13	\\
080123	& Late	& 0.50\phantom{0}    & 53.63$\pm$7.67	& -20.89	& 10.12	& 0.43	& 0.29	& ---	& ---	& $4.16_{-1.81}^{+2.32}$	& 2,13	\\
140622A	& Late	& 0.96\phantom{0}    & 32.95$\pm$11.25	& -21.80	& 10.17	& 0.66	& 0.24	& ---	& ---	& $4.24_{-1.85}^{+2.35}$	& 2,13	\\
160408A	& Late	& 1.91$^\star$       & 14.13$\pm$1.25	& -20.88	& 9.32	& 0.62	& 1.84	& ---	& ---	& $2.20_{-0.99}^{+1.61}$	& 2,13	\\
170728A	& Late	& 1.49\phantom{0}    & 32.25$\pm$3.01	& -19.85	& 10.09	& 0.16	& 2.07	& ---	& ---	& $3.48_{-1.49}^{+2.07}$	& 2,13	\\
200411A	& Late	& 0.83$^\star$       & 41.98$\pm$7.48	& -21.53	& 10.23	& 0.62	& 2.99	& ---	& ---	& $4.33_{-1.89}^{+2.38}$	& 2,13	\\
201221D	& Late	& 1.06\phantom{0}    & 29.35$\pm$24.09	& -20.96	& 9.36	& 0.27	& 0.10	& ---	& ---	& $2.48_{-1.11}^{+1.69}$	& 2,13	\\
210919A	& Late	& 0.24\phantom{0}    & 51.05$\pm$1.92	& -19.42	& 9.87	& 1.62	& 0.32	& ---	& ---	& $4.19_{-1.82}^{+2.30}$	& 2,13	\\
        \noalign{\smallskip}
        \hline
	\end{tabular}
    \egroup
	\begin{flushleft}
        \textbf{Notes.} 
        $^\star$ Photometric redshift inferred from SED fitting by \cite{nugent2022}. 
        $^\S$ Sersic profile of the dominant component.\\ 
        \textbf{References.} 
        (1) \citet{fong2010};
	    (2) \citet{fong2022};
	    (3) \citet{berger2007};
	    (4) \citet{fong2013};
	    (5) \citet{fong2016};
	    (6) \citet{blanchard2017};
	    (7) \citet{rouco2021};
	    (8) \citet{paterson2020};
	    (9) \citet{fong2021};
	    (10) \citet{rastinejad2022};
	    (11) \citet{nugent2020};
	    (12) \citet{oconnor2021};
	    (13) \citet{nugent2022}.
	\end{flushleft}
\end{table*}

\subsection{Galactic potentials}

The first step in modelling the trajectories of SGRB progenitors is to model the gravitational potential of the host galaxies. 
As the host sample is heterogeneous in terms of galaxy properties, our goal is to build a model that is general enough to be suitable for all hosts, and yet detailed enough to reproduce the potentials to a good approximation. 
To this end, we start from the method of \cite{church2011} who modelled the dark halo potentials for a similar galaxy sample, and we build on it mainly by including the stellar potential and the star formation history (SFH). 
We assume the potentials to remain constant with time, and discuss the implications of this assumption in Sect.~\ref{sec:3.2}.

\subsubsection{Dark matter potential}

Following \cite{church2011}, we model the dark matter halo with a logarithmic potential 
\begin{equation}
    \rho_\mathrm{dm}(r) = \frac{v_\mathrm{h}^2}{4\pi G} \frac{3r_\mathrm{h}^2+r^2}{(r_\mathrm{h}^2+r^2)^2}
\end{equation}
where $r_\mathrm{h}$ is the core radius of the halo and $v_\mathrm{h}$ is the circular velocity at infinity. 
These two parameters are derived host-by-host from the $B$-band absolute magnitude $M_B$ using scaling relations. 
For early-type hosts, we use the relations of \cite{thomas2009}, while for late-type hosts, we use the relations of \cite{kormendy2016}. 
We obtain the magnitudes $M_B$ from the SED models fitted by \cite{nugent2022}, by integrating them over the $B$-band filter in the rest-frame wavelengths.

\subsubsection{Stellar light distribution}\label{sec:2.2.2}

To model the stellar light distribution, we start from the S{\'e}rsic models \citep[see \citealt{graham2005} for a concise overview]{sersic1963,sersic1968} for the surface brightness $\Sigma$ of the host galaxies, which are in the form
\begin{equation}
    \Sigma(R) = \Sigma_0\, \exp \left[ -b_n \left( \frac{R}{R_\mathrm{half}} \right)^{1/n} \right]
\end{equation}
where $R$ is the projected radius from the galaxy centre, $\Sigma_0$ is the central surface brightness, $n$ is the S{\'e}rsic index, $R_\mathrm{half}$ is the half-light radius, and $b_n$ is a function of $n$ approximated as
\begin{equation}
    b_n = 2n -\frac{1}{3} +\frac{4}{405n}+\frac{46}{25515n^2}
\end{equation}
following \cite{ciottibertin1999}.
We collect the values of $R_\mathrm{half}$ and $n$ from the literature and list them in Table~\ref{tab:1}. 
In a few cases the surface brightness is fitted with a combination of two S{\'e}rsic profiles. 
For these galaxies we only report the dominant component in Table~\ref{tab:1}, but we do include both components weighted by $\Sigma_0$ in our modelling.

We deproject the surface brightness into the light density distribution according to one of two assumptions, which are meant to reproduce the two extremes. 
We either assume that all the light is concentrated in an infinitely thin disc, in which case its radial distribution coincides with the S{\'e}rsic profile, or we assume that light has a spherically-symmetric density distribution, in which case we need to actually deproject the S{\'e}rsic profile. 
Hereafter, we refer to the distributions from these two models as "discs" and "spheroids", respectively.

For the spheroids, we deproject the S{\'e}rsic profiles into the light density $\varrho$ using the approximation of \cite{vitral2020}
\begin{equation}
    \label{eq:4}
    \varrho(r,n) = \varrho_\mathrm{PS}(r,n)\ \mathrm{dex} \left[ \sum^{10}_{i=0}\sum^{10}_{i=0} a_{ij} \log^i r \log^j n \right]
\end{equation}
which improves that of \cite{ps1997}
\begin{equation}
    \varrho_\mathrm{PS}(r) = \varrho_0\, \left( \frac{r}{R_\mathrm{half}} \right)^{-p_n} \exp \left[ -b_\mathrm{n} \left( \frac{r}{R_\mathrm{half}} \right)^{1/n} \right].
\end{equation}
Here, $p_n$ is a function of $n$ for which we use the improved approximation of \cite{lgm1999}
\begin{equation}
    p_n = 1 -\frac{0.6097}{n} +\frac{0.05463}{n^2}
\end{equation}
instead of the original one from \cite{ps1997}, while for the coefficients $a_{ij}$ we use the values tabulated in Table~B.2 of \cite{vitral2020}.

Since we have S{\'e}rsic profiles for only 20 out of 70 hosts, we also employ a second method to estimate the stellar light distribution that can be applied to the whole sample. 
For $R_\mathrm{half}$, we estimate its value from the mass-size distribution of \cite{vdw2014} using the stellar mass $M_\star$ given by the SED fits. 
We employ their mass-size distributions rather than parametric fits because they better reproduce $R_\mathrm{half}$ when extrapolated at $M_\star\lesssim 10^{9}\ \mathrm{M}_\odot$ \citep[c.f.][]{nedkova2021}. 
For $n$ instead, we simply assume $n=1$ for the discs \citep[which gives an exponential disc profile;][]{freeman1970} and $n=4$ for the spheroids \citep[which gives a de Vaucouleurs profile typical of spheroids and elliptical galaxies;[]{dv1948}.

\subsubsection{Stellar potential}

For discs, we model the stellar potential with a double-exponential disc of density
\begin{equation}
    \label{eq:7}
    \rho_\star(R,z) = \rho_{\star,0}\, \exp \left( -\frac{R}{h_R}-\frac{|z|}{h_z} \right)
\end{equation}
where the scale length is $h_R = R_\mathrm{half}/b_1$ and the scale height is assumed to be $h_z=\gamma\, h_R$.
We adopted a fixed ratio of $\gamma=0.2$ given that the observed value for local galaxies ranges between 0.1 and 0.3 \citep{padilla2008,unterborn2008,rodriguez2013}.
The disc potential is implemented using the approximation with three Miyamoto-Nagai discs of \citet[originally introduced by \citealt{flynn1996}]{smith2015}.

For spheroids, we assume that mass follows light and model the stellar potential from the deprojected light distribution we get with Eq.~\ref{eq:4} using the self-consistent field method of \cite{hernquist1992}. 
For both spheroids and discs we compute the normalisation factor $\rho_{\star,0}$ so that the total mass is equal to the stellar mass $M_\star$ given by the SED fits.

\subsection{Stellar population models}

\begin{figure*}
    \centering
    \includegraphics[scale=1]{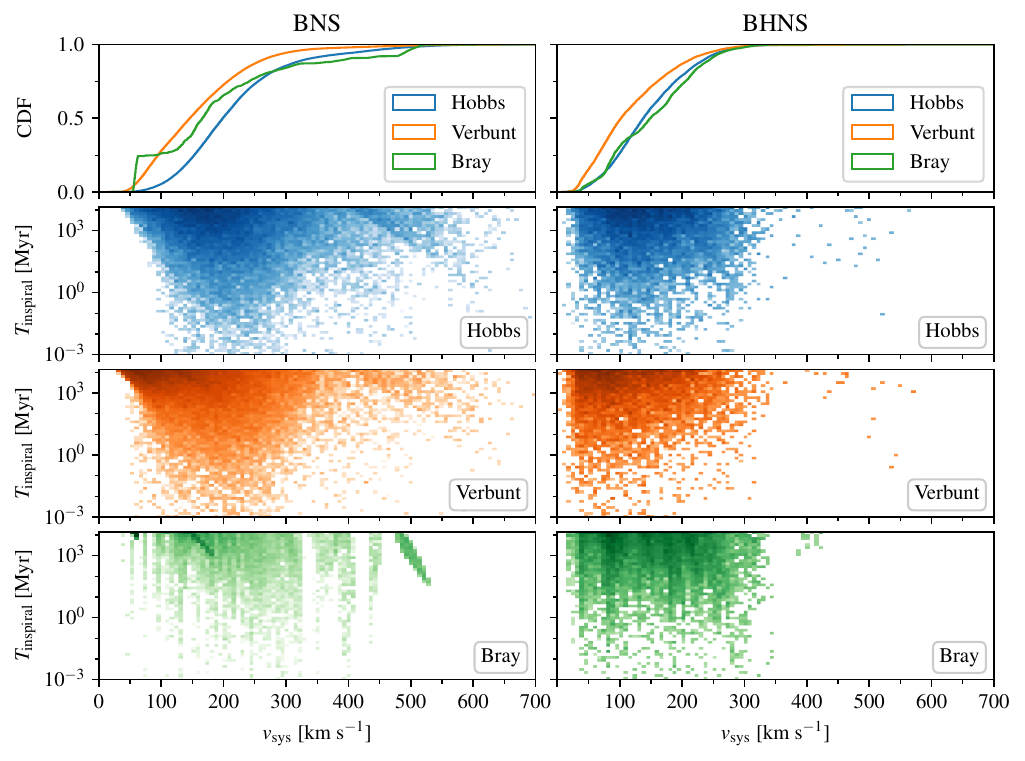}
    \caption{Merger times $T_\mathrm{inspiral}$ and systemic kicks from the second SN $v_\mathrm{sys}$ for BNSs and NSBH binaries from our population synthesis models. The populations are obtained using three different models for the SN natal kicks, namely the Hobbs model \citep{hobbs2005}, the Verbunt model \citep{verbunt2017}, and the Bray model \citep{bray2016}. Distributions from the Bray model present discrete features because the natal kicks from this prescription are weighted by the compact remnant mass output by BPASS, which in turn reflects the discrete grid of input parameters (e.g. initial masses, orbital periods).}
    \label{fig:2}
\end{figure*}

We make use of the Binary Population and Spectral Synthesis \citep[BPASS v2.2.1;][]{eldridge2017,stanway2018} binary stellar evolution models. These are a suite of models across binary parameter space that implement detailed modelling of binary interactions. The relative frequency of the models are weighted according to the initial binary parameters including primary mass, mass ratio and orbital period following \cite{moe2017}. A \cite{kroupa2001} initial mass function with slope -1.30 below 0.5\msol and -2.35 above (up to 300\msol) is adopted. The final weightings correspond to the number of each binary system expected, at zero-age main sequence (ZAMS), in a stellar population of $10^{6}\msol$. Models at \about half Solar metallicity ($Z=0.010$) are adopted for this work.

We determine the systemic velocities and delay times of BNS and NSBH binaries as follows. Core-collapse is deemed to occur when a star, at the end of the model, has a total mass >1.5\msol, a carbon-oxygen core mass >1.38\msol, and a non-zero oxygen-neon core mass. Remnant masses are pre-calculated as a \texttt{BPASS} output following \cite{eldridge2004}. If the remnant has a mass in the range $1.38 < M/\msol < 3.0$ we declare the object to be a neutron star, more massive remnants are black holes. The ejected mass is simply the difference between the pre-SN and remnant masses \citep[for a detailed study of alternative remnant mass prescriptions in \texttt{BPASS} see][]{briel2023}. We determine if the binary stays bound - and if so, what the properties of the new orbit are - using the model of \citet[][see also \citealt{tauris1999}]{tauris1998}. This also determines the systemic velocities of the binaries which remain bound. For neutron star natal kicks we adopt three distributions: the single-peaked Hobbs distribution \citep{hobbs2005}, the double-peaked Verbunt distribution \citep{verbunt2017} and the Bray model \citep[which ties the natal kick magnitude to the remnant and ejecta mass]{bray2016,richards2023}. Remnant natal kick magnitudes are drawn from these distributions, and the directions are sampled isotropically. For black holes, the kick magnitude is reduced by a factor of 1.4\msol/$M$ \citep{eldridge2017} to account for lower black hole natal kick velocities \citep[e.g.][]{mandel2016,atri2019}.

All orbits that survive the first SN are circularised and the radius is set equal to the semi-latus rectum $a(1-e^2)$, where $a$ and $e$ are the semi-major axis and eccentricity output by the \citet{tauris1998} model, while we leave the orbit eccentric after the second SN. For models which end with a bound compact binary, we compute the gravitational wave in-spiral time following \citet{mandel2021}.

We therefore have, for each model which ultimately ends with a compact binary, the times between ZAMS, first SN, second SN, and merger, and the systemic kicks after each SN.
From the above, we can determine for any given model - when seeded in a galactic potential - when and where the final BNS or NSBH binary will merge, with respect to the birth time and place. In Fig.~\ref{fig:2}, we show the time between second SN and merger $T_\mathrm{inspiral}$ and the systemic kick $v_\mathrm{sys}$ from the second SN for all the models we employ.

\subsection{Galactic trajectories}

The BNS and NSBH merger locations are modelled by seeding the \texttt{BPASS} populations within the host potential and simulating their galactic trajectories with \texttt{galpy}\footnote{\url{https://github.com/jobovy/galpy}} \citep{bovy2015}. The potentials are built by summing the dark halo potential and the stellar potential, the first of which is modelled using the routine \texttt{potential.LogarithmicHaloPotential}, and the second using either \texttt{potential.MN3ExponentialDiskPotential} (for discs) or \texttt{potential.SCFPotential} (for spheroids).

The binaries are seeded in space following the stellar light as expected for core-collapse SNe \citep{fruchter2006}, specifically by sampling their initial galactocentric radii from the light density distribution with the inverse transform method. We then compute the circular velocities at the binary location in each specific potential, and initialise the binaries on circular orbits. The orbital planes are oriented in random isotropic directions, in order for the models to be agnostic with respect to the galaxy viewing angle. To model the SFH instead, we weight each binary model \emph{a posteriori} with the normalised star formation rate at the lookback time that coincides with their delay time (namely, the time between ZAMS and merger).

After seeding, we apply the systemic kick predicted for the first SN by \texttt{BPASS}, simulate the galactic trajectory up to the second SN, apply the second systemic kick, and then simulate the trajectory up to the merger. Both kicks are imparted with a direction that is randomly sampled from the isotropic distribution. The resulting offsets are then projected on a random isotropic orientation for comparison with the observed ones.


\section{Results}\label{sec:3}

\subsection{Comparing predicted offsets to observations}\label{sec:3.1}

\begin{figure*}
    \centering
    \includegraphics[scale=1]{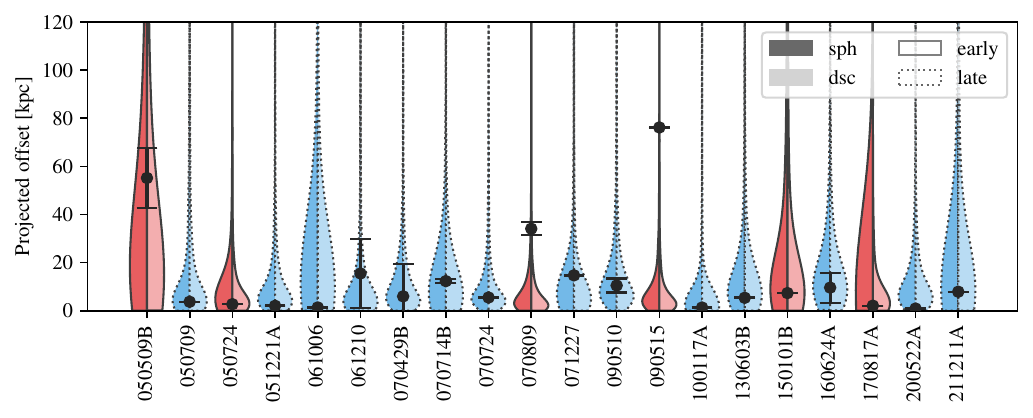}
    \caption{Predicted and observed merger offsets of BNSs for the subsample of SGRBs in the BRIGHT catalogue \citep{fong2022,nugent2022} that have a S\'{e}rsic fit. The violins show the distributions of predicted offsets either assuming discs (dsc) or spheroids (sph) for the stellar light distribution. The dots show the observed SGRB offsets with their respective errors. We also indicate whether the host galaxy is early- or late-type.}
    \label{fig:3}
\end{figure*}

\begin{figure*}
    \centering
    \includegraphics[scale=1]{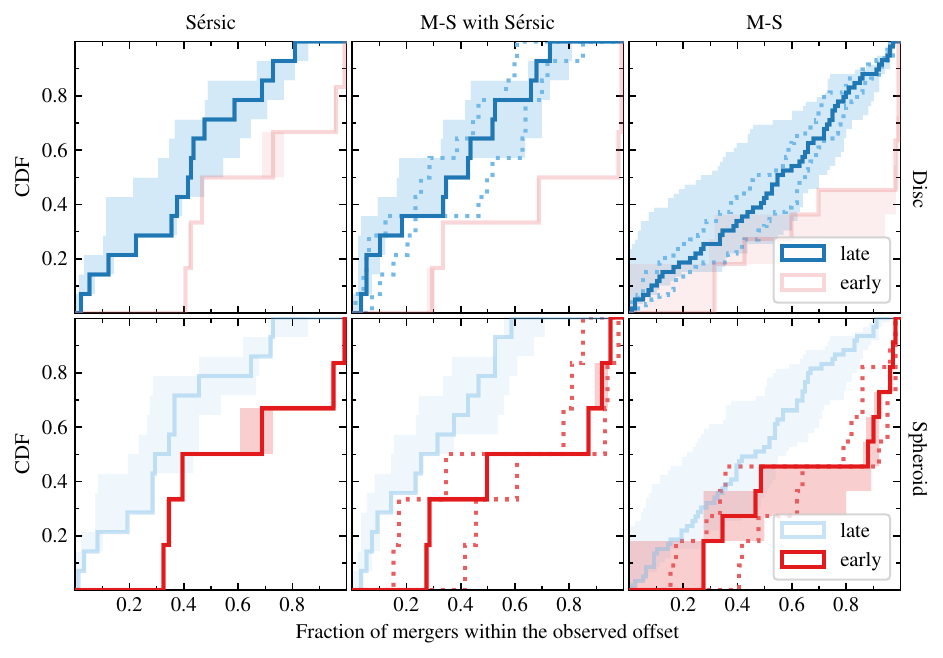}
    \caption{Cumulative distributions (CDFs) of the fraction of BNS mergers within the observed offset $F$ for the fiducial model. Top row is from from models where the stellar component is a disc, while the bottom row is for models with a spheroid. Left column is from the hosts with a S\'{e}rsic fit, central column is from the hosts with a S\'{e}rsic fit but modelled with the mass-size relation, right column is from all the hosts modelled with the mass-size relation. The shaded areas indicates the uncertainty due to the astrometric error on the observed offset, while the dotted lines indicates the uncertainty due to the spread of the mass-size relation. The host types are shown in the legends.}
    \label{fig:4}
\end{figure*}

From our models, we can predict the distribution of merger offsets for each host galaxy and for a range of assumptions, the first of which is whether the stellar component is a disc or a spheroid. 
Hereafter, we will assume that the fiducial models are discs for late-type hosts and spheroids for early-type hosts.
In Fig.~\ref{fig:3} we show the predicted offsets for the subsample that has a S\'{e}rsic fit, with both disc and spheroid models. Second, we have three different models for the SN natal kick. We assume the Verbunt model as fiducial and provide a comparison to the other two in Sect.~\ref{sec:3.2}. 

A third assumption we make is whether the stellar light distribution is modelled from a S\'{e}rsic fit or the mass-size relation of \citet[][see Sect.~\ref{sec:2.2.2}]{vdw2014}. Since the hosts with a S\'{e}rsic fit account for around 1/4 of the whole sample, and hence modelling stellar light through the mass-size relation allows us to analyse a significantly larger sample, we decide to not adopt one of the two assumptions as fiducial. 

To compare predictions to observations, we rely on the fraction $F$ of predicted mergers having a projected offset smaller than the observed one. We have therefore one realisation of $F$ for each host with a value between 0 and 1, with 0 meaning that all predicted offsets are larger than the observed one, and 1 meaning that all predicted mergers happen within the observed offset. If we were to sample one value of $F$ for each host according to the predicted offset distributions, $F$ would then be uniformly distributed between 0 and 1. We can therefore compare predictions to observations by testing the $F$ distribution against the uniform distribution with a Kolmogorov-Smirnov (KS) test.

In Fig.~\ref{fig:4} we show the $F$ distribution for a range of assumptions with the fiducial models highlighted. In particular, we show the distributions obtained using the S\'{e}rsic fit, the mass-size relation, and the mass-size relation applied only to the subsample with a S\'{e}rsic fit. The latter (shown in the middle column in Fig.~\ref{fig:4}) allows us to check whether using the mass-size relation changes the $F$ distribution significantly compared to the more accurate S\'{e}rsic models. We notice that the fiducial distribution for late-types do not change significantly, while for early-types we have 2 points out of 6 that move to slightly higher values. In the same figure we also show the uncertanties due to errors on the observed offsets (which combine astrometric errors from the merger locations and the galaxy centroids) and the uncertanties due to the mass-size relation spread. We notice that for early-types the offset errors dominate over the mass-size relation spread, while for early-types they have comparable magnitudes.

\subsection{Impact of the natal kick model}\label{sec:3.2}

\begin{figure*}
    \centering
    \includegraphics[scale=1]{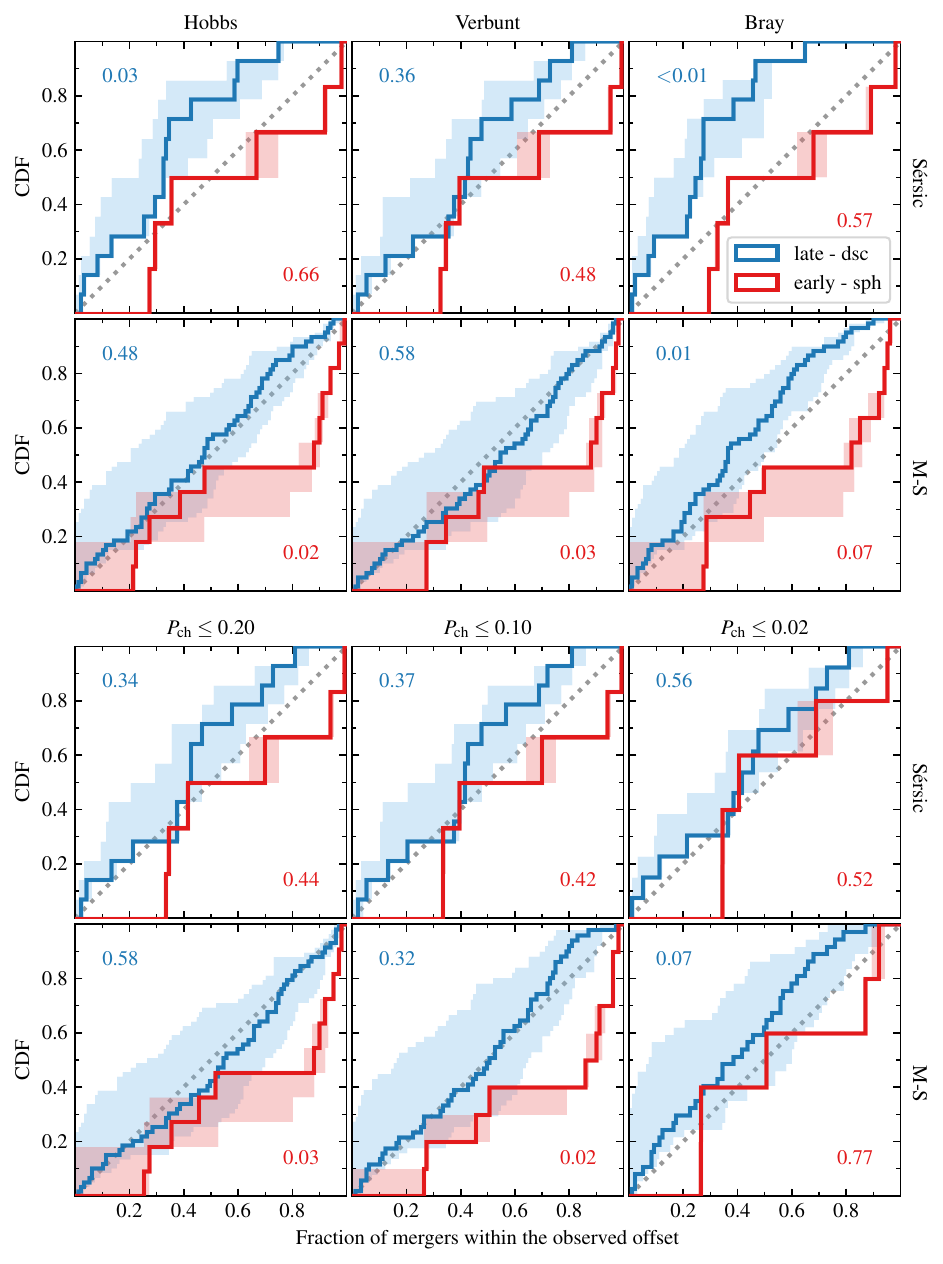}
    \caption{Cumulative distributions (CDFs) of the fraction of BNS mergers within the observed offset $F$. \emph{Upper panels}. Comparison between models with different natal kick prescriptions, namely the Hobbs model (left column), the Verbunt model (central column), and the Bray model (right column). Distributions in the first row are modelled with S\'{e}rsic profiles, while those in the second row are modelled with the mass-size relation. The KS $p$-value for each CDFs with respect to a uniform distributions is shown in each panel. Host types and stellar distribution models are shown in the legend. \emph{Lower panels}. Comparison between hosts subsamples at different $P_\mathrm{ch}$ tresholds, namely $\leq 20\%$ (left column), $\leq 10\%$ (central column), and $\leq 2\%$ (right column). All the remaining assumptions are fiducial, in particular we are using the Verbunt model for natal kicks.}
    \label{fig:5}
\end{figure*}

When comparing $F$ distributions to the uniform distribution in Fig.~\ref{fig:5}, we note that a cumulative distribution function (CDF) above the bisector (which is the CDF of the uniform distribution) indicates that our model is underestimating the observed distribution of merger offsets, while a CDF below the bisector indicates that we are overestimating the offsets. From Fig.~\ref{fig:5} we find that the overall trend is for the CDFs from late-type hosts to be systematically above those from early-type hosts. This can be the result of several effects, either alone or in combination, such as the scaling relations predicting systematically lighter dark halos for late-type hosts, intrinsically larger offsets in early-type hosts, or a bias against large offsets for late-type from the host association criterium (see Sect.~\ref{sec:3.3} for the latter).

Looking at the subsample with S\'{e}rsic fits (first row in Fig.~\ref{fig:5}), we see that the null hypothesis can be rejected at a significance level below 5\% for Hobbs and Bray kicks since they overestimate the offset distributions of late-type hosts, while no kick model can be rejected based on the offsets of early-type hosts. 
This is consistent with our previous work \citep{gaspari2024b} showing that shallower potentials such as those of spiral or dwarf galaxies can discriminate between kick models, whereas deeper potentials like those of massive ellipticals cannot. We also note that our analysis is probing the upper part of the kick distribution. Indeed, while the kick distribution from the Hobbs model is altogether shifted toward higher values compared to the Verbunt model, the Bray kicks peak in between those from Hobbs and Verbunt, but with a noticeable excess of high kicks (see Fig.~\ref{fig:2}) which is what results in binaries migrating and merging further outward in the potential. The overestimate of merger offsets in late-type galaxies could also be explained by the sample being contaminated with collapsar-driven GRBs (e.g. traditional long GRBs), which are concentrated on the host light in contrast to the offset SGRBs \citep{fong2013,blanchard2016,fong2022}. However, \citet{fong2022} find that the possible contaminants (which they identify using the \citealt{bromberg2013} criterium) and the remaining population have undistinguishable offset distributions, therefore making this explanation unlikely. Overall, when comparing the results for late-type offsets from the subsample modelled with S\'{e}rsic to those from the full sample modelled with the mass-size relation, we see that a larger sample is warranted in order to do model selection. 

Looking at the sample modelled with the mass-size relation (second row in Fig.~\ref{fig:5}), we see that only Bray kicks are rejected by late-type hosts, while all models but Bray are rejected by early-type hosts (though $p$-value for Bray is 7\%). Furthermore, the $F$ distributions of early-type show a remarkable bimodality that is independent of kicks and is inconsistent with a uniform distribution, having around half of the sample at $F<0.5$ and the remaining half at $F>0.8$.
This excess of high $F$ values can be the result of either galaxy evolution, the progenitor population being more extended than we estimate, spurious host associations (see Sect.~\ref{sec:3.3} for the latter), or any combination of them. By comparing the observed SGRB offsets to the BNS merger offsets of \citet{wiggins2018}, \citet{fong2022} show that the fraction of population missing at large offests ($\geq 30-50 kpc$) from the BRIGHT catalogue should not be substantial, and it is consistent with the fraction of inconclusive host associations ($\sim 7\%$). If this were the case, the excess at high $F$ could be even more pronounced.

Starting from the first explanation, namely galaxy evolution, we note that our models assume the galactic potentials to be static, in contrast to real galaxies which grow in mass and size over cosmic time \citep[e.g.][]{vdw2014} and especially the most massive ones whose evolution can be dominated by galaxy mergers \citep[e.g.][]{rodriguezgomez2015,rodriguezgomez2016}.
To this regard, \citet{kelley2010} and \citet{wiggins2018} have shown that models accounting for the time evolution of the host potential and its neighbours predict larger BNS merger offsets, except for the BNSs that had the lowest systemic kicks and those merging at high redshifts. 
The early-type hosts in our sample have indeed an old stellar population and redshifts $z\lesssim 0.5$, supporting the galaxy evolution scenario, whereas late-type hosts tend to be much younger and span a redshift range of $0\leq z \lesssim 2.5$ \citep[see Fig.~8 of][]{nugent2022}.
On top of the evolving potential, in an environment with a high number density of galaxies there is also the mixing of neighbouring populations \citep[at least $\sim5-13\%$ of SGRBs hosts belong to a galaxy cluster,][]{nugent2020}. \citet{zemp2009} showed that in these dense environments, the central galaxy in the most massive halos retains BNSs better than those in the field, but the BNSs escaping from their satellites produce much more diffuse distribution mergers that might still be associated with the central galaxy instead of the one where they originated from. In this case, the association with the massive ellitpical galaxy may be spurious, but the association with the larger structure in which it resides may be correct. 

For the second explanation, namely that the progenitor population is more extended than we estimate, we identify three possibilities. First, it might be that the mass-size relation predicts smaller half-light radii than the true ones. This would result in the BNSs being seeded deeper in the potential with a higher initial circular velocity, which in turn would dampen the migration outward. We do not find strong evidence to either support or discard this scenario, although we notice that if this were the case then it would indicate a selection effect in the population of early-type hosts. Second, our sample might include SGRBs produced by the merger of BNS dynamically formed in dense environment such as globular clusters (GCs). \citet{church2011} estimate that $\leq 10\%$ of SGRBs might originate from this channel, however more recent simulations showed that the merger rate of dynamically-formed BNSs is negligible compared to field binaries \citep[][c.f. \citealt{grindlay2006}]{ye2020}, hence we find this scenario unlikely. A third and last possibility is that early-type hosts have a significant fraction of stellar mass in a highly extended halo which has a surface brightness below the detection limit, and hence it is undetected in conventional observations (compared to stacked images or ultra-deep observations). In this scenario BNSs can be born already with a large normalised offset and merge \emph{in situ}, though this would require no systemic kicks \citep{perets2021}. We notice that this is in tension with the observational evidence supporting systemic kicks \citep{atri2019,zhao2023,odoherty2023,disberg2024} and models showing that large merger offsets can also be achieved with modest kicks \citep[$v_\mathrm{sys}\lesssim100$ \kms;][]{gaspari2024a,disberg2024}, and therefore we find this scenario also unlikely. 

For comparison, we show the results obtained using NSBH binaries instead of BNSs in Fig.~\ref{fig:A.1}. Here we see that the $F$ distributions from all natal kick models are consistent with the uniform distribution when using the subsample with S\'{e}rsic fits, but 5 out of 6 fail the KS test when using the whole sample modelled with the mass-size relation. In particular, we see that for the latter sample the $F$ distributions are always below the bisector in the case of late-type hosts, which is consistent with NSBH binaries having smaller systemic kicks (Fig.~\ref{fig:2}) and thus predicted merger offsets that are systematically smaller than the observed ones. This again supports our finding that a larger sample is warranted for model selection. We also notice that $F$ distributions for early-type hosts are bimodal, as for BNSs.

\subsection{Impact of the host association criterium}\label{sec:3.3}

To understand the impact of possible mis-identified host galaxies, we repeat the analysis separately for each of the three $P_\mathrm{ch}$ cuts, namely $P_\mathrm{ch}\leq20\%$ (Gold, Silver and Bronze hosts), $P_\mathrm{ch}\leq10\%$ (Gold and Silver hosts), and $P_\mathrm{ch}\leq2\%$ (Gold hosts only). These three samples contain respectively 70, 62, and 42 host galaxies, and we expect they have respectively $\leq4.4$, $\leq2.8$ and $\leq0.8$ mis-identified hosts.

The probabilty of chance alignment $P_\mathrm{ch}$ is known to be biased against hosts that are faint and have small apparent sizes, either because they are intrinsically faint or at high redshift \citep{levan2007,berger2010,tunnicliffe2013,oconnor2022}. The reason for this is twofold. First is because $P_\mathrm{ch}$ takes into account the transient localisation error and the host apparent size in such a way that faint hosts can have low $P_\mathrm{ch}$ only if the transient is well localised (e.g. has sub-arcsecond localisation from an optical/infrared counterpart) and is nearby in projection, in contrast to a bright or extended host which can maintain a low $P_\mathrm{ch}$ even with a worse transient localisation or a greater offsets \citep[e.g.][]{gaspari2024b}. Second, intrinsically faint galaxies have shallower potentials \citep{kormendy2016}, hence BNSs can more easily escape the potential and merge at greater offsets, where $P_\mathrm{ch}$ is close to unity \citep[e.g.][]{gaspari2024b}.

Looking at the subsample with S\'{e}rsic fits (third row in Fig.~\ref{fig:5}), we notice no significant difference in the $F$ distributions when applying different $P_\mathrm{ch}$ cuts. However, when looking at the sample modelled with the mass-size relation (fourth row in Fig.~\ref{fig:5}), we do see the $F$ distributions of late-type hosts moving leftward of the bisector as we cut the sample at lower $P_\mathrm{ch}$, while for early-type hosts there is no evident difference and the bimodality is still present at all cuts. 

The shift above the bisector in the case of late-type hosts is consistent with the exclusion of the galaxies at the largest offsets from the subsamples with the lower $P_\mathrm{ch}$ cuts, and the subsequent systematic underestimate of predicted offsets with respect to the observed ones. \citet{fong2022} shows that the inclusion of Silver and Bronze hosts capture a substantial number of bursts at $z\gtrsim 1$, of hosts with lower luminosities ($\leq 10^{10}$ L$_\sun$ in the $r$-band), and hosts at larger offsets (with the median offset increasing by $\sim 3$ kpc), thus diversifying the population of known SGRB hosts and resulting in a more representative sample \citep[see also][]{nugent2022}. Our result further supports this claim.

Moving to early-type hosts, the bimodality of $F$ at all $P_\mathrm{ch}$ thresholds suggests that this feature is not the result of spurious host associations. Indeed, as the early-type subsample contains 5 Gold ($P_\mathrm{ch}\leq2\%$) and 6 Silver ($P_\mathrm{ch}\leq10\%$) hosts, we already expected $\leq0.7$ mis-identified hosts ($\leq6.4\%$). This supports the scenario in which bimodality is produced by shortcomings in our models, first and foremost the inability to capture the complex effects of galaxy evolution and structure formation. However, we are not fully confident in ruling out a substantial contribution from mis-identified hosts at large offsets given the biased nature of $P_\mathrm{ch}$, as explained at the beginning of this Section.


\section{Conclusions}\label{sec:4}

In this work we predicted the galactocentric offset of 70 observed SGRBs from the BRIGHT catalogue \citep{fong2022,nugent2022}, by modelling the galactic potentials on a host-by-host basis and seeding in them synthetic populations of BNSs and NSBH binaries from the BPASS code \citep{eldridge2017,stanway2018}. The host sample is divided into three subsamples based on the chance alignment probability $P_\mathrm{ch}$ (42 Gold hosts with $P_\mathrm{ch}\leq 2\%$, 20 Silver hosts with $2\%<P_\mathrm{ch}\leq 10\%$, 8 Bronze hosts with $10\%<P_\mathrm{ch}\leq 20\%$), however we use the whole sample for the fiducial models. We reproduced each galactic potential by summing the potential of a dark halo to that of the stellar component, the first obtained from the galaxy total magnitude in the $B$-band through scaling relations \citep{thomas2009,kormendy2016}, and the second obtained by deprojecting the galaxy surface brightness profile normalised to the total stellar mass inferred from SED fitting \citep{nugent2022}. As fiducial models, we assumed the stellar component of late-type hosts (i.e. star-forming) to be a disc, and that of early-type hosts (i.e. quiescent and transitioning) to be a spheroid. The synthetic binaries are seeded in the potentials using stellar light as a proxy, and in time using the SFH fitted by \citet{nugent2022}. We then simulated the galactic trajectories with \texttt{galpy} \citep{bovy2015} accounting for the velocity kicks received at each SN, and recorded the merger location for analysis. In the population synthesis, we employed three different natal kick prescriptions \citep[namely][]{hobbs2005,bray2016,verbunt2017} to probe the impact of SN kicks on the offsets, taking the Verbunt model as fiducial. Since we use static galactic potentials, our models cannot reproduce the time evolution of the host potentials and its possible neighbours, as has been done by theoretical models that employ cosmological simulations of structure formation \citep[e.g.][]{zemp2009,kelley2010,behroozi2014,wiggins2018}. However, the strength of our methodology is that it can be applied to an observed transient population using the observed galaxy properties. 
We summarise our conclusions:

\begin{enumerate}

\item Regarding late-type hosts, we find that our fiducial model for BNS mergers is consistent with observed SGRB offsets (last two rows of Fig.~\ref{fig:5}). However, our fiducial sample includes hosts with $P_\mathrm{ch}$ up to 20\%, in contrast to the more common criterium for a strong host association which is $P_\mathrm{ch}$ below a few per cent. When we cut our host sample at lower $P_\mathrm{ch}$ thresholds, our predictions tend to increasingly overestimate the offsets, which is consistent with the exclusion of hosts at the largest offsets. 
Therefore, we conclude that a less conservative $P_\mathrm{ch}$ threshold should be considered when associating late-type hosts, given that our predictions for BNS merger offsets are consistent with SGRB offsets up to $P_\mathrm{ch}\leq 20\%$.

\item Regarding early-type hosts, we find that our models significantly underestimate the offsets of around half the subsample (last two rows of Fig.~\ref{fig:5}). We find this result regardless of the $P_\mathrm{ch}$ threshold adopted, therefore the discrepancy may not be caused by mis-identified hosts although we notice that association by $P_{ch}$ might introduce a bias toward bright and extended galaxies. We conclude that there are two plausible explanations (see Sect.~\ref{sec:3.3}). First, our models lack the temporal evolution of the host potential and its possible neighbours, which would spread out the spatial distribution of mergers. Second, in the presence of satellite galaxies, the distribution of mergers originating from the satellites can overlap that of the associate host, and produce a spurious association due to mixing of neighbouring populations.

\item When comparing the predictions for BNS mergers from different natal kicks models, we do not find strong evidence against any of them, as they all produce similar results (first two rows of Fig.~\ref{fig:5}). When we turn to NSBH binaries however, our models sistematically underestimate the observed offsets (first two rows of Fig.~\ref{fig:A.1}), likely due to the lower systemic kicks we predict. Also, when comparing results from the subsample of hosts with S\'{e}rsic fits to those from the whole sample, we see that a larger and more diverse sample is warranted for model selection. 

\end{enumerate}

Our results strongly support the origin of the at least the majority of SGRBs in the merger of kicked compact objects, and suggest that there is not significant contamination from, for example, collapsar GRBs. Although our results are inconclusive regarding the natal kick prescriptions we tested, we point out that systemic kicks are not only determined by natal kicks, but also by the pre-SN orbit and the progenitor masses through the mass-loss kick. Thus, systemic kicks also encode the evolutionary processes of the progenitor binary, such as mass transfer and common envelope episodes, and might be correlated to merger times. For these reasons, a natural follow-up of this work is to test the main free parameters and uncertainties in population synthesis modelling, and we suggest that merger offsets get included in future model selection as an additional observational constraint along the currently used statistics such as merger rates.


\begin{acknowledgements}
We thank Ross Church, Michela Mapelli, Rosalba Perna, Maria Celeste Artale, Ilya Mandel, and Wen-fai Fong for the useful discussions.
NG acknowledges studentship support from the Dutch Research Council (NWO) under the project number 680.92.18.02.
AJL was supported by the European Research Council (ERC) under the European Union’s Horizon 2020 research and innovation programme (grant agreement No.~725246). 
AAC acknowledges support through the European Space Agency (ESA) research fellowship programme.
\end{acknowledgements}

\bibliographystyle{aa}
\bibliography{biblio}


\begin{appendix}

\section{Merger offsets of NSBH binaries}\label{sec:A.1}

\begin{figure*}
    \centering
    \includegraphics[scale=1]{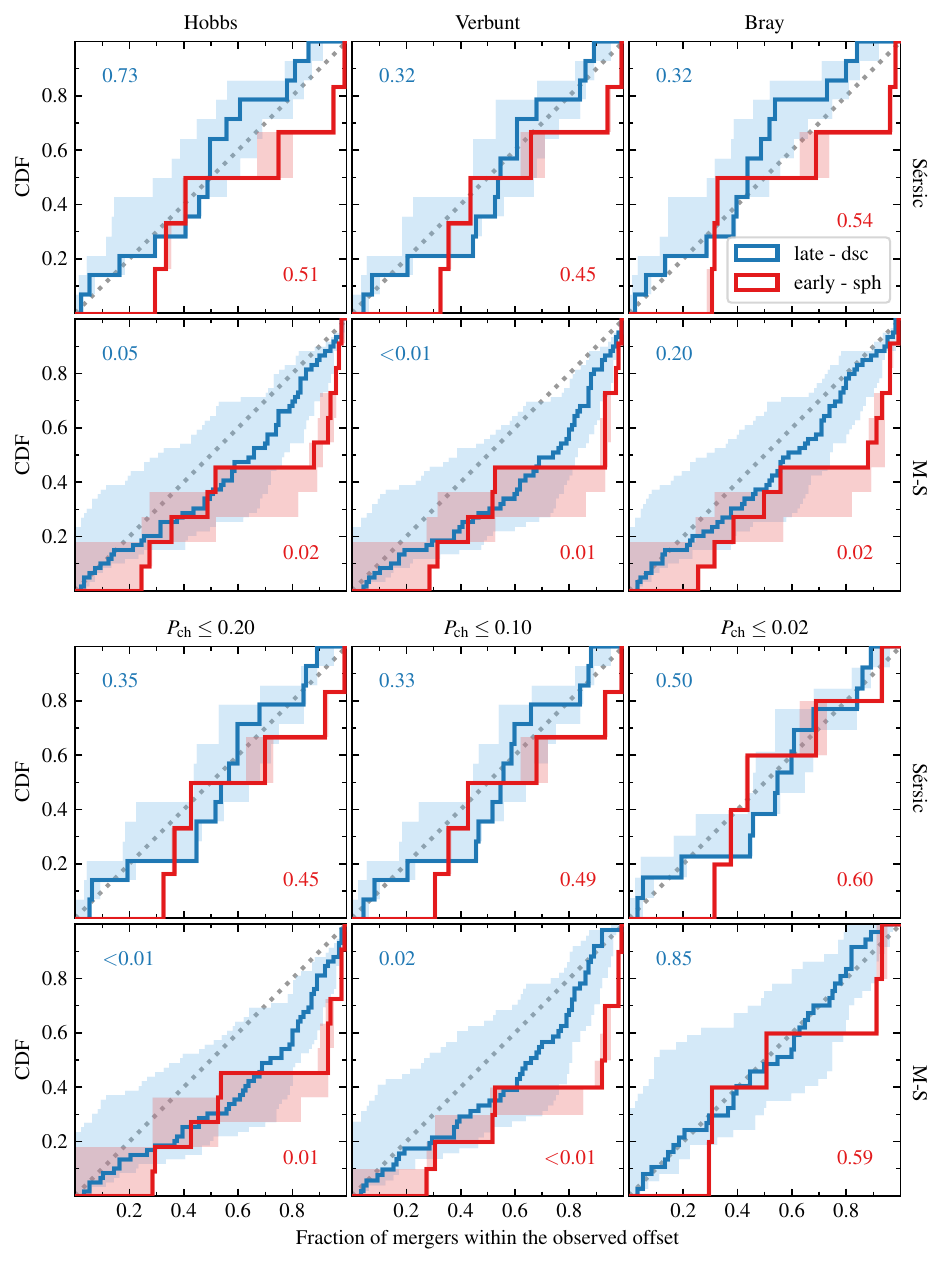}
    \caption{Same as Fig.~\ref{fig:5}, but for NSBH mergers.}
    \label{fig:A.1}
\end{figure*}

\end{appendix}

\end{document}